\documentclass[conference]{IEEEtran}

\usepackage[pdftex]{graphicx}
\usepackage{amsmath}
\usepackage{amssymb}
\usepackage{eqparbox}
\usepackage[nolist]{acronym}
\usepackage{colortbl}
\usepackage{enumitem}
\usepackage{caption}
\begin{document}

\title{\LARGE Evaluation of Ultra Wide Band Technology as an Enhancement for BLE Based Location Estimation}
\author{\authorblockN{Miro Roman Bilge\\miro.bilge(at)gmail.com\\Stuttgart Media University}
\authorblockA{Stuttgart, Germany}}

\maketitle


\begin{abstract}
In the scope of this paper, the precise positioning of objects with the help of Ultra Wide Band technology is evaluated. 
To achieve this, a prototype module for the use with the Decawave transceiver DW1000 was implemented and added to FruityMesh, an \ac{IoT} mesh network of the company M-Way Solutions.
The focus of the network is low-power communication based on \acf{BLE}. This includes, for example, Building Automation, Lighting Management and Asset Tracking. Especially the last stated point benefits from a position tracking as precise as possible and a long battery life. Therefore, FruityMesh is used to perform a signal-strength based localization with the existing \ac{BLE} messages.
Due to absorption, interference, and diffraction, these measurements tend to fluctuate and allow a positioning accuracy within $\pm$1.5m. With the integration of Ultra Wide Band and the \acf{ToA} method, a centimeter-precise localization was made possible in the course of this work, which at the same time causes only small additional costs in general. Since the connection of extra hardware is associated with decreased energy efficiency, an algorithm for optimizing the control was first developed and then tested against the created scenarios. In addition to the motion-based control of the hardware, various configurations and adjustments were analyzed to reduce the power consumption by \ac{UWB} transmission.
Finally, the developed prototype was compared with a realistic reproduction of the existing Asset Tracking to evaluate the benefits for the use in the productive application.
\end{abstract}



\section{Introduction}

The ongoing relevance of localizing objects and people is further underscored with the upcoming of social distancing. While tracking systems such as the \ac{GPS} perform well outdoors, they do not provide the required accuracy or even any results inside buildings. In addition to the absence of the direct path, walls and ceilings result in signal attenuation. Several approaches for indoor positioning exist for established technologies such as Wi-Fi or \ac{BLE}, which are already present in facilities such as office environments or warehouses. 
For example, the Asset Tracking of the \ac{IoT} platform BlueRange developed by the company M-Way Solutions. This indoor tracking is based on the \ac{IoT} firmware FruityMesh, which is an energy-efficient mesh network built on \ac{BLE}. With the ability of battery operation, FruityMesh as part of BlueRange can be used for communication, building automation, as well as lighting control \cite{bluerange}. 
When an asset tag is moved within an existing BlueRange infrastructure, its position is detectable with an accuracy of $\pm1.5m$. 
With low cost and a long lasting battery, BlueRange Asset Tracking offers the opportunity for applications such as equipment inventory or intelligent utilization of rooms and machines. 
For usage scenarios that require more precise and accurate localization, especially in situations without \ac{LOS}, the existing tracking can not provide the required accuracy. Also, the extension of the infrastructure by another technology is reviewed in this work.
To be relevant for the existing BlueRange Asset Tracking, the following points have to be considered in the examination:
\begin{itemize}[topsep=5pt] \label{list:fruityreq}
\item \textbf{Applicability} - Energy efficiency for battery operation.
\item \textbf{Scalability} - Multiple deployment without interference.
\item \textbf{Accuracy} - Improved indoor localization.
\item \textbf{Responsiveness} - High report interval.
\item \textbf{Coverage} - Supported range for distance estimation. 
\item \textbf{Cost \& Complexity} - Comparable price without infrastructure changes.
\end{itemize}

\subsection{State of the Art}

Many of the solutions and technologies considered for extending the BlueRange Asset Tracking are heavily used in our daily life. Paying by card is done with \ac{RFID} or infrared is used to detect movement for switching on the lights. With properties like low-cost and low-power, these two technologies offer a promising starting point for improving the current tracking method. On the basis of \ac{AoA}, the attachment of an infrared sensor to a shopping cart allows its localization with an error level of $\pm$0.6m \cite{InfraredCart}. In addition, infrared offers the advantage of high data throughput of up to 12.5Gb/s \cite{infrareddata}\cite[9]{zafari2019survey}.

Further, promising approaches are presented in the articles on long-range as well as \ac{UHF} \ac{RFID} \cite{rfidlr}\cite{passrfid}\cite{rfid} and are separated into active and passive methods. While the battery powered long-range tag consumes 21.3$\mu W$ to provide active distance estimation up to 35m, passive \ac{RFID} tags are completely supplied via induction. In this process, the energy for response transmission is obtained from the received electromagnetic signal. Thus, even without a battery installed, the phase of the \ac{RFID} signals allows position estimation up to 4m with a median error of 10cm \cite{rfidCart}. 

In contrast, \ac{sonar}, as a passive approach of the Ultrasound technology, uses the reflections of its emitted signals for the ranging. Based on the same principle, recent studies managed to track an object in a test environment by recording its echoes \cite{ultrasound}. 
In order to identify the actual echoes of the object, the experimental setup requires a previous calibration. 
Subsequently, the \ac{ToF} of the echoes allows the distance calculation with an average error of $\pm$0.105m.

By combining individual techniques, for example for the clock synchronization of the participants, the measurement results can be refined. In this experiment \cite{infrasync}, ultrasound and \ac{IR} are used as a hybrid system achieving an accuracy of a few centimeters. While distances up to 5m are estimated with ultrasound, the synchronization is kept with \ac{IR}.

In the search of an improvement for the existing \ac{BLE} Asset Tracking, the extension by new technologies is suggested.
However, BLE provides enhancements with its latest standard 5.1 too. 
Direction finding with the \ac{AoA} offers a new procedure for positioning, but also requires hardware extension. Nevertheless, this technique is compared to the previous presented methods limited in coverage and accuracy \cite{deadonarrival}.

Therefore, this analysis focuses on positive properties to complement \ac{BLE}. Here, \ac{UWB} offers not only centimeter-accurate measurements, but also excellent multipath performance due to its wide bandwidth. Described in these studies \cite{concurrent}\cite{uwblocal}\cite{CompareRadio}, the precise time resolution allows an accurate distance estimation over the \ac{ToF}. In addition to the equipment of the asset tags, the introduced technologies also require an upgrade to the infrastructure. 

In that scenario, Wi-Fi has the advantage that it is widely available in office and work environments and therefore the acquisition costs for Wi-Fi-based systems is low. While offering a range ten times higher than \ac{BLE}, Wi-Fi offers great advantages, but is negatively affected from interference in the \ac{ISM} bands \cite[7,8]{zafari2019survey}. 

Especially indoor measurement using \ac{RSSI}, \ac{ToF} or \ac{AoA} tend to fluctuate \cite{iBeacon}. 
However, the approaches in \cite{Wideep} and \cite{DeepFi} show that distance estimation with existing Wi-Fi installations is possible, but requires post processing for a mean error below 2m.

Filtering the possible solutions by the defined requirements quickly clears the picture.
The power consumption of systems based on Ultrasound or the Wi-Fi versions 802.11\textit{a/b/g/n} is too high to be powered by battery, like the FruityMesh asset tags. The coin cells, used to power the asset tags up to one year, would not last an hour providing Wi-Fi distance estimation \cite{concurrent}. The same applies to the \ac{IR} integration in FruityMesh. While this technique offers a lower price and a higher accuracy, \ac{NLOS} as well as sunlight results in poor measurement quality.
The \ac{RFID} technology is promising, but due to sampling times of 2s, dynamic scenarios can only be partially covered \cite{rfidCart}. 
Additionally, indoor measurements are negatively affected by interference \cite{rfidlr}.

In comparison, \ac{UWB} requires only sub-nanoseconds for the measurement and offers a high accuracy with low power requirements \cite{dw1000datasheet}. Especially in technology-rich indoor environments with numerous obstacles and multipath signals, the technology provides a strong distinction of the ranging signals compared to narrow band transmission \cite[338,340]{bensky}. 

In general, the \acl{UWB} technology fits the requirements and promises to improve the BlueRange Asset Tracking. With great properties in the absence of the direct path and centimeter accuracy, this approach creates the possibility to track devices in any size. 
In order to verify these positive qualities and to be able to assess negative aspects, this work evaluates the integration of UWB technology for the BlueRange network. In the process, the theoretical principles were fully implemented in practice in form of a prototype. By initiating, distributing and evaluating the UWB measurement results over the mesh network, the integration can be confirmed as successful while maintaining the existing BlueRange functionalities. Thereby, the excellent measurement characteristics can be approved in LOS as well as in NLOS scenarios. While UWB tracking offers significantly higher accuracy over a longer distance, the energy characteristics do not meet comparable values, despite intelligent control and low-power UWB configuration.


\section{Methodology}

In the course of evaluating UWB Technology as an extension of BlueRange Asset Tracking, it was first necessary to examine the technology itself.

Since the restriction of the \ac{UWB} regulation is limited to transmission strength and frequencies, the physical and the software layer remain free in their definition. Whilst providing a high degree of flexibility, it causes the development to start at a hardware-related level.

Therefore, suitable \ac{UWB} configurations focusing on energy-efficiency as well as on high measurement accuracy are defined \cite[32]{Sahinoglu}\cite{802.15.4-2011}.
To prove the functionality of the theoretical considerations and to analyze the actual performance, test scenarios in an actual environment are designed and executed. For this reason, a suitable distance estimation method based on \ac{UWB} was implemented and compared with a simplified replication of the existing signal-strength based approach. 
In this study, a prototype integrating \ac{UWB}-capable hardware as part of the BlueRange network was developed.
This provides the distribution of distance measurement results over the mesh network and possible future productive use.
Since the FruityMesh microcontroller does not support \ac{UWB} by default, the hardware had to be extended for this purpose. This is associated with additional costs and increased power consumption. As a result, a concept focusing the energy-efficient control had to be implemented for the operation of the \ac{UWB} hardware.
To limit the scope of the project and focus on the \ac{UWB} integration, a mobile application replaced the server and gateway of the BlueRange network. With the development of the application, an interface for analysis and storage of the data was created. 
Furthermore, local restrictions were removed and testing in different scenarios including outdoor installations was realized. 

Finally, based on theoretical studies of the \ac{UWB} technology and the used hardware, an appropriate configuration for the \ac{UWB} communication had to be chosen and tested with various application tests.
By analyzing the advantages and disadvantages in terms of accuracy and energy efficiency, the conclusion was drawn whether \ac{UWB} can improve the existing Asset Tracking approach. 
\\
Thus, the development of the prototype for evaluation could be divided into three parts:
\begin{itemize} \label{list:methods}
\item The UWB hardware configuration, with the goal of achieving high quality measurements combined with low power consumption.
\item The definition and implementation of the UWB message exchange for ranging.
\item A concept for the control and integration into the FruityMesh firmware and thus the BlueRange network.
\end{itemize}

\subsection{Theoretical Background}
Regardless of the comparative recent interest in UWB, the origin of the technology dates back to over a hundred years. Using Heinrich Hertz's experiments as an inspiration, Guglielmo Marconi managed to transmit data with UWB pulses. Due to the focus on narrow band transmission, the high potential of UWB with data rates above 500Mb/s was largely overseen. Despite several patents that were applied in the 70's, no commercial interest arose due to a lack of regulation. In 2002, after studies about the low impact of UWB on other technologies, such as \ac{GPS}, had been published, a regulation by the \ac{FCC} finally took place \cite{Stroh2003WidebandMU}\cite[3]{10.5555/1841062}. 

In general, \ac{UWB} technology is a method to establish near-field communication, using radio signals that have at least a bandwidth of 500MHz or 20\% of their fractional bandwidth \cite[20]{Sahinoglu}. As an overlay system, the energy of the signal is distributed over a wide frequency spectrum, overlapping frequency bands that are already in use by other applications and services. Due to the wide dispersion, the power spectrum of the \acs{UWB} signal is so small that it is lost in the background noise of narrow-band transmission \cite[332]{bensky}. 
To not interfere with other narrow band technologies, the \ac{UWB} signals are only allowed to use the bandwidth from 3.1 to 10.6GHz. This range is further limited regarding power density, installation of antennas and regulated differently for each country worldwide \cite{AllgZTUwb}.

At present, the idea of accurate and precise Asset Tracking based on \ac{UWB} is adopted by a number of companies.
Several integrable \acp{IC}, such as beSpoon MOD1, DecaWave DW1000 and NXP SR040, are considered for use. \cite{decawave_2019}\cite{bespoon_2021}\cite{nxp_2020}. 
Since the NXP SR040 is not available at the time of this work and has not been tested yet in a wide variety of environments, it cannot be used despite promising consumption properties \cite{nxp_2020}. Moreover, the price for the expected quantity is 2.5 times lower by using the Decawave \ac{IC} \cite{symmetryelectronics}.
In addition, the beSpoon MOD1 and the Decawave DW1000 have been tested for years. Especially the \ac{UWB} module of Decawave offers many advantages with its forum, numerous examples and a wide distribution in commercial solutions \cite{decawave_2019}\cite{bespoon_2021}. 
Decawave's DW1001 development board is shipped with their UWB module and a nRF52832 chip, which is supported by the FruityMesh firmware.
This ensures a fully mesh-enabled prototype without the need to design and commission a custom board \cite{dw1001datasheet}. 
Thus, sources of error caused by the hardware design can be excluded and the calibration parameters as well as the default configuration of Decawave can be used as a basis.

As a result of the technology assessment and due to all the benefits described above, the Decawave chip was chosen for the prototype and the technology comparison in this work.

\subsection{Platform}
\begin{figure}[ht!] 
\centering
\caption{\ac{UWB} Transmission Frame}
\label{fig:frame}
\includegraphics[width=3.45in]{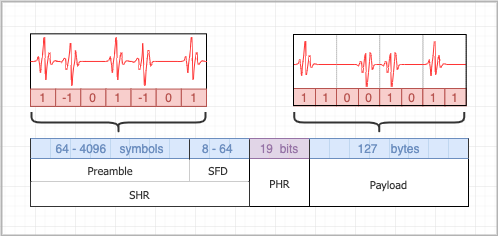}
\end{figure}

While Marconi managed to transmit messages in Morse code with simple pulses, modulation techniques are used to achieve higher data rates and reliable performance \cite[4]{10.5555/1841062}. This legacy method is called \ac{IR-UWB} and uses sequences of narrow pulses for data transmission. 
The different modulation techniques are combined in the UWB frame, shown in figure \ref{fig:frame}, which encapsulates user data for transmission.

The front part of the frame, the \ac{SHR}, is used to detect and synchronize the \ac{UWB} messages. For better recognition, the preamble as well as the \ac{SFD} are formed by symbols of the ternary alphabet. The '1' and '0' of \ac{OOK} are supplemented by an inverted pulse for '-1' \cite[212]{dw1000userman}. 
20 different preamble codes are available, which determine the order of the symbols. The choice of the preamble code, adapted to the \ac{PRF} and the selected bandwidth, is not handled by the hardware and must be set by software. Apart from this, the implementation of UWB transmission methods also requires a selection among different preamble lengths. 
Supported in a range of 64 to 4096 symbols by the DW1000 module, a long preamble improves the \ac{SNR}, while a short one with less transmission time consumes less power.
In addition, the DW1000 offers the option of defining custom sequences and thus, improving the reception. 
However, transmitting with user configured \ac{SFD} codes breaks the compatibility to standard \ac{UWB} hardware \cite[213]{dw1000userman}\cite[233-235]{mobilepositioning}. 

The \ac{PHR} and the payload are both modulated with \ac{BPM-BPSK}. Using this combination, two bits can be represented in one symbol, as shown in figure \ref{fig:frame}. BPM maps one bit over the shifted position of the burst. Another bit is expressed by the amplitude of the burst through \ac{BPSK}. The \ac{PHR} needs 19 of these bits containing information about the \ac{UWB} message. Especially the included and by the developer definable data rate as well as the length of the payload are essential for the prototype. With data rates of 110kb/s, 850kb/s and 6.8Mb/s and up to 1023 bytes of user data, the selected setup is decisive for the power consumption \cite{dw1000datasheet},\cite[28,211]{dw1000userman}.

Essential for the choice of a suitable transmission configuration is the selection of a channel and the \ac{PRF}. Defined by the center frequency, the DW1000 module supports 6 channels from 3.5 to 6.5GHz. 
In general, channels with a lower frequency require less power and have a longer range due to lower attenuation. This can be calculated with the \ac{FSPL} equation of Friis \cite{aps06}. 
Another factor affecting the transmission of signals is the \ac{PRF}. It specifies the number of signals that can be transmitted per second. Supported are 16 and 64MHz. With 24.04ns more transmission time per preamble symbol, the PRF of 64Mhz has a marginally higher power consumption \cite[212]{dw1000userman}\cite[17]{dw1000datasheet}.

As used in this work, the DW1000 module from Decawave offers many options to define the \ac{UWB} transmission itself. Thus, in compliance with hardware technical requirements, an own UWB ranging procedure can be developed focusing on minimal power consumption. 
For this purpose, the different configuration options have to be evaluated and the relevant components for the UWB messages have to be defined. While Decawave's settings bring device-specific advantages, following the IEEE standard can simplify future integration with other systems and products. However, due to the integration in BlueRange and the presence of a BLE mesh network, there are further potential opportunities for UWB Asset Tracking.

Based on the theoretical principles, six configuration modes for \ac{UWB} transmission are developed in the course of this work (See table \ref{tab:configs}). 

\begin{table}[ht!] 
\centering
\caption{Theoretically developed \ac{UWB} Test Configurations}
\label{tab:configs}
\begin{tabular}{|l|l|l|l|l|l|} 
\hline
\rowcolor[rgb]{0.855,0.91,0.988} Mode & Data Rate & \begin{tabular}[c]{@{}>{\cellcolor[rgb]{0.855,0.91,0.988}}l@{}}PRF \\(MHz) \end{tabular} & \begin{tabular}[c]{@{}>{\cellcolor[rgb]{0.855,0.91,0.988}}l@{}}Preamble \\Length + PAC\end{tabular} & CH & Attributes \\ 
\hline
1 & 6.8 Mbps & 16 & 64 + 8~ & 1 & Min. Power \\ 
\hline
\rowcolor[rgb]{0.937,0.937,0.937} 2 & 6.8 Mbps & 16 & 128 + 8 & 1 & \\ 
\hline
3 & 6.8 Mbps & 64 & 128 + 8 & 1 & \\ 
\hline
\rowcolor[rgb]{0.937,0.937,0.937} 4 & 6.8 Mbps & 64 & 128 + 8~ & 5 & \\ 
\hline
5 & 110 kbps & 64 & 4096 + 64~ & 1 & Max. Range \\ 
\hline
\rowcolor[rgb]{0.937,0.937,0.937} 6 & 6.8 Mbps & 64 & 128 + 8 & 4 & Max. BW \\
\hline
\end{tabular}
\end{table}

Beginning with the low-power mode, which focuses on a minimal power consumption by utilizing the smallest supported frequency and the shortest transmission time. Due to poor signal quality, such as low \ac{SNR}, the measurement results can suffer in accuracy and precision. Therefore, the signal quality is gradually increased at the expense of energy efficiency. 
Additionally, mode 5 and 6 were used to determine the maximum range for the transmission with the DW1000 and to analyze the influence of a higher bandwidth.


The determination of a suitable configuration and the resulting UWB transmission are the basis for indoor localization. For this purpose, a method for distance estimation must be defined and implemented first. Afterwards, the position can be determined by lateration. This requires at least three fixed anchor nodes in a three-dimensional space. 
Simplified, the asset tag is located at the intersection of the measured distances to the respective anchor nodes. 

In addition, the current BlueRange Asset Tracking operates with distance estimates based on \ac{BLE} signal strength as well. Thus, similar procedures and methods can be used for post-processing.
While the \ac{UWB} messages also provide the \ac{RSSI} value for the distance calculation using the \ac{FSPL} equation, the \ac{ToA} allows a more precise measurement method \cite[148,152]{bensky}\cite[14]{Sahinoglu}. 

Since \ac{UWB} signals propagate in free space at the speed of light, the distance can be estimated by the \ac{ToF} \cite[72]{kuepper}. Furthermore, the short UWB signals provide excellent multipath characteristics and are easily distinguishable from each other. This allows precise time measurement, which is required at speeds of 3.3ns/m. 
Every time a \ac{UWB} message is transmitted or received by the DW1000 module, a timestamp is generated. Using these \ac{RX} and \ac{TX} timestamps, the ToF can then be calculated.
\begin{figure}[ht!]
 \centering
 \centering
 \caption{\acf{SS-TWR}}
 \label{fig:sstwr}
 \includegraphics[width=3.1in]{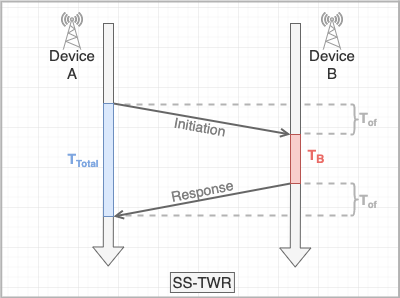} 
\end{figure}

Shown in figure \ref{fig:sstwr}, the first \ac{TX} timestamp is created by device A, while device B records the first \ac{RX} timestamp.
In this process, a clock drift cannot be avoided when using multiple DW1000 modules. This means that each chip's clock oscillates with a difference of only a few parts per million. The influence is, nevertheless, so immense that such a measurement result would not be reliable \cite[6]{dw1001datasheet}. 

Therefore, one possibility to solve this would be the synchronization of all anchor nodes. Defining one master clock, the remaining units would be supplied with the clock signal by a wired connection. This takes all the flexibility out of battery-powered Asset Tracking based on a decentralized mesh network. Another solution and the method used in the prototype of this study is \ac{TWR}. Here, a \ac{UWB} message completes at least one round trip between two nodes. Afterwards, the clock drift error can be corrected almost completely \cite{aps11}.

In order to minimize transmission time and the associated power consumption during this process, 
the \ac{SS-TWR} (See figure \ref{fig:sstwr}) method is used, in which only one message is transmitted back and forth \cite{concurrent}. 
\\Furthermore, message parts of the \ac{IEEE} standard and the Decawave protocol were removed to shorten the message length to a minimum.
In the course of this process, unnecessary fields for the prototype, such as the sequence number, were removed and the redundant part in the timestamps was excluded. This trimmed the ranging response to 12 bytes, which is only half the length used by Decawave's approach. 
This simplification is viable, because the \ac{UWB} prototype is based on the FruityMesh firmware, which is coordinating the individual nodes and the data exchange. 

Thus, with this extension by a second technology for improving indoor localization, the power consumption is increased in any case. 
While the \ac{UWB} configuration and the message length affects the energy required by the transmission, the \textit{Idle} state, in which the DW1000 module is ready for transmission, can also be optimized. 

For this purpose, the DW1000 module supports two different sleep modes with 50nA and 1$\mu A$ power consumption. 
To achieve a similar battery life as the existing BlueRange asset tags, the more energy-efficient \textit{Deep-Sleep} mode is initiated as soon as the \ac{UWB} asset tag movement ends.
However, this mode can only be quit externally, which is handled by the nRF52 chip. This FruityMesh controller also manages the wake-up process in case of motion, which is detected by the installed accelerometer. 
As part of the \ac{SS-TWR}, the fixed anchor nodes are put into \textit{Deep-Sleep} state after receiving the last \ac{UWB} message and waiting for a certain offset $T_{\text{Offset}}$. During this process, the existing BlueRange functions, such as receiving the asset tags advertising messages, continue to run. Depending on the motion value in these BLE messages, the Anchor node detects moving asset tags in its nearby environment and the UWB module can be switched on again. 
Further, the anchor node can inform additional nodes through the FruityMesh network and increase the range $N_{\text{Range}}$ of the \ac{UWB} distance measurement. 

Parameters such as $N_{\text{Range}}$ , $T_{\text{Offset}}$ or the position update frequency $F_{\text{Update}}$ vary depending on local conditions and requirements. They must be adjusted based on battery size, number of asset tags, and size of the network. In addition, it must be clarified whether the accuracy and precision of one distance measurement is sufficient or whether several measurements must be performed, as in the BLE approach. Consequently, various tests were carried out to determine and classify these values. 


\section{Results and Discussion}

The specified configurations (See table \ref{tab:configs}) were analyzed in terms of time duration and power consumption. For this purpose, the Nordic \ac{PPK} was used to determine the actual time periods and the drawn current of a \ac{SS-TWR} performed at 10m distance. 
\begin{table}[ht!] 
\centering
 \caption{Power Measurements - Length (ms)}
\label{tab:ppktime}
\begin{tabular}{|l|l|l|l|l|l|l|} 
\hline
\rowcolor[rgb]{0.855,0.91,0.988} Mode & $\sum$ & TX & RX & Idle & Init & $ \sigma $ \\ 
\hline
1& 22.391 & 0.910 & 13.820 & 4.284 & 3.377 & $ \pm 0.13 $ \\ 
\hline
\rowcolor[rgb]{0.937,0.937,0.937} 2 & 22.775 & 1.037 & 14.075 & 4.284 & 3.378 & $ \pm 0.13 $ \\ 
\hline
 3 & 22.772 & 1.038 & 14.074 & 4.282 & 3.377 & $ \pm 0.13 $ \\ 
\hline
\rowcolor[rgb]{0.937,0.937,0.937}4& 22.772 & 1.038 & 14.076 & 4.283 & 3.375 & $ \pm 0.13 $ \\ 
\hline
5& 36.020 & 5.254 & 22.909 & 4.281 & 3.376 & $ \pm 0.13 $ \\ 
\hline
\rowcolor[rgb]{0.937,0.937,0.937} 6 & 22.768 & 1.037 & 14.075 & 4.279 & 3.375 & $ \pm 0.13 $ \\
\hline
\end{tabular}
\end{table}

As shown in table \ref{tab:ppktime}, the \ac{UWB} transmission consists of the parts \textit{\ac{TX}}, \textit{\ac{RX}}, \textit{Idle} and the initialization. The initialization is the time the DW1000 module needs to wake up and stabilize the crystal oscillator. Afterwards, the nRF host controller writes the data to be sent to the DW1000, which forms the \textit{Idle} part. Finally, the actual transmission begins. Here, the \ac{RX} time is significantly longer than the \ac{TX} time. This is caused by the reception, the processing and the response of the second anchor node. Although this process of the anchor node is kept short with efficient programming and high-priority interrupts, it has a large influence with approximately 10ms. 
To calculate the \ac{ToF}, this duration is estimated and written to the response message before its actual transmission. Therefore, it has been chosen rather high to avoid a timeout on the anchor side. Even if other program functions are being executed, this delay ensures a response within the time limit. If the host controller is always on time, it can be further optimized. 

Similarly, the asset tag will not receive a response before the anchors processing time expires, thus its reception can be delayed as well. For this purpose, the theoretical \ac{SS-TWR} duration can be calculated and compared with the measured data in the table \ref{tab:ppktime}. With a difference of 11.58ms, the asset tag can be switched to \ac{RX} state at least 10ms later without fluctuations causing errors.

Looking at the remaining times, it is noticeable that the two non-variable states \textit{Idle} and \textit{Init} also have a large impact on the total time. While the duration of the initialization is fixed due to the crystal oscillator, the \textit{Idle} time can be reduced a bit further by fetching the registries Clock PLL flag. This can be achieved by rewriting the wake up method provided by Decawaves \ac{API}. Since the PLL flag depends on the used channel \cite[22]{dw1000userman}, it is not changed to keep comparability between the defined configurations.

\begin{table}[ht!]
\centering
\caption{Power Measurements - Average Current (mA)}
\label{tab:ppkcurrent}
\begin{tabular}{|l|l|l|l|l|l|} 
\hline
\rowcolor[rgb]{0.855,0.91,0.988} M. & $\varnothing$ & TX & RX & Idle & Init \\ 
\hline
1 & $ 53$ & $ 25.0 \pm 0.8 $ & $ 76.9\pm 4.1$ & $ 17.4 \pm 1.1 $ & $ 9.1 \pm0.8 $ \\ 
\hline
\rowcolor[rgb]{0.937,0.937,0.937} 2 & $ 56$ & $ 30.7 \pm 1.2 $ & $ 79.2 \pm 4.2$ & $ 18.9 \pm 0.9 $ & $9.0 \pm 0.8 $ \\ 
\hline
3 & $ 59$ & $ 32.5 \pm 1.2 $ & $ 85.0 \pm 4.2$ & $ 18.6 \pm 0.5 $ & $ 9.0 \pm 0.8 $ \\ 
\hline
\rowcolor[rgb]{0.937,0.937,0.937} 4 & $ 64$ & $ 33.4 \pm 1.1 $ & $ 91.9 \pm 4.1$ & $ 18.2 \pm 0.9 $ & $ 9.1 \pm 0.8 $ \\ 
\hline
5 & $ 63$ & $ 33.4 \pm 1.1 $ & $ 85.2 \pm 4.2$ & $ 17.3 \pm 0.5 $ & $ 9.0 \pm 0.8 $ \\ 
\hline
\rowcolor[rgb]{0.937,0.937,0.937} 6 & $ 70$ & $ 41.5 \pm 4.0 $ & $ 101.5 \pm 4.0$ & $ 17.3 \pm 0.9 $ & $ 9.0 \pm 1.1 $ \\
\hline
\end{tabular}
\end{table}

Nevertheless, the average times demonstrate that by using a high data rate and a short message length the transmission time as well as the power consumption can be kept low. 
Small changes, such as increasing the preamble length from 64 to 128 bytes in configuration 2, do not have a significant effect on the total time.
This results from the large period for response processing, the \textit{Idle} and the \textit{Init} state, which are not affected by the various configurations.
Thus, depending on the transmission energy, this adjustment can be used to improve the quality of the measurement. 
\\
Therefore, the actually required power was analyzed in the next step. 
While the DW1000 module consumes on average only 50nA in the \textit{Deep-Sleep} state, the complete \ac{UWB} asset tag still requires around 850$\mu A$. This increased basic power consumption arises from the activation of the mesh functions, the debug and the SEGGER interface. With these functions, the configuration can be easily changed via the developed mobile application without reflashing the \ac{SoC}. 
By deactivating this functions for productive use, a marginally improved energy efficiency is achieved \cite{dw1000datasheet}.
Depending on the part of the UWB transmission, an increased current demand is additionally measured, as shown in table \ref{tab:ppkcurrent}.

In general, the start of the \ac{SS-TWR} results in a 60 times higher average power consumption.
Depending on the selected parameters, sending and receiving \ac{UWB} messages requires almost 32\% less energy. Implementing the 16MHz \ac{PRF} option, for example, reduces the power consumption by around 5-11\%. Based on the measured average consumption, channel 1 with the lowest center frequency and a \ac{PRF} of 16MHz should be selected for an energy-efficient application.

\subsection{Accuracy \& Precision Measurements}

In order to determine the accuracy and precision for indoor localization, the implemented distance estimation was executed over a range of 1-30m in an office hallway. For this purpose, two UWB asset tags were placed at \ac{LOS} and the same height. Subsequently, 1000 cycles of SS-TWR were performed at each of ten specified distances up to 30m. In order to prevent the measurement from being influenced by the heating of the module, only one measurement per second was processed and the temperature was monitored.
To minimize external influences, the results were received by a mobile application over the FruityMesh network, just as in the productive use. In addition to persisting and exporting data, this mobile cross-platform app was developed to offer the basic functions of the other BlueRange components. Furthermore, the app can be used to change the prototype configuration for testing and to adjust other settings without the need to intervene in the setup.

Despite these ideal conditions, a productive system with a preamble length of 64 symbols is hardly feasible. While all modes with a longer preamble length achieve a reception rate of almost 100\%, the lowest power configuration fails. Only 17.78\% of the performed rankings were completed successfully. Thus, the asset tag would have to initiate up to six distance estimations per anchor node to obtain a usable measurement result. This lack of reliability increases the transmission time and consumes more energy. Since only the strongest signals were received and used for distance estimation, the promising average standard deviation of only 4.36cm is not significant. Based on this, mode 1 can not be used for a productive system and is excluded from any further evaluation and the following \ac{LOS} outdoor test. 

The other configurations with a data rate of 6.8Mb/s allow transmission over distances of 109 to 124m.
With the successful execution of half of all \acp{TWR} at a distance of 622m, the configuration 5 outperforms the other modes in terms of range. In addition to utilizing a low density of nodes, this excellent range also offers new possibilities for the FruityMesh network. When used as a data link, for example, different networks can be connected over long distances. For indoor tracking applications, the range of around 100m is absolutely sufficient to cover the available distances in office buildings or warehouses. Consequently, all remaining configurations are suitable.
\begin{figure}[ht!]
 \centering
 \centering
 \caption{SS-TWR - Range Bias 0-30m - Mode 4}
 \label{fig:evabias}
 \includegraphics[width=3.45in]{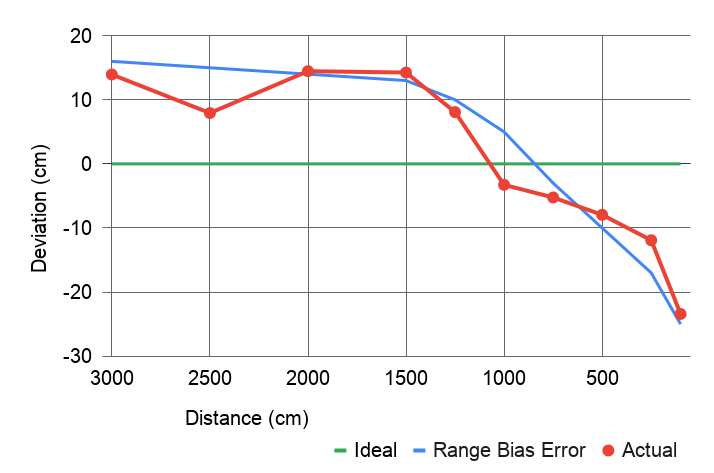} 
\end{figure}

The deviation of the short range measurements is negative and changes to positive for the larger test distances.
This characteristic is caused by the range bias error and is related to the signal strength. As seen in the Diagram \ref{fig:evabias}, the deviation below 5m is up to -25cm. Within the range of 5 to 10m the error is mainly below 10cm. After this range, the reported results increase to +15cm.
Due to this continuity, the error can be calculated and corrected by adding a correction factor.
\begin{figure}[ht!]
 \centering
 \centering
 \caption{SS-TWR - Deviation at 250cm}
 \label{fig:evacandle}
 \includegraphics[width=3.4in]{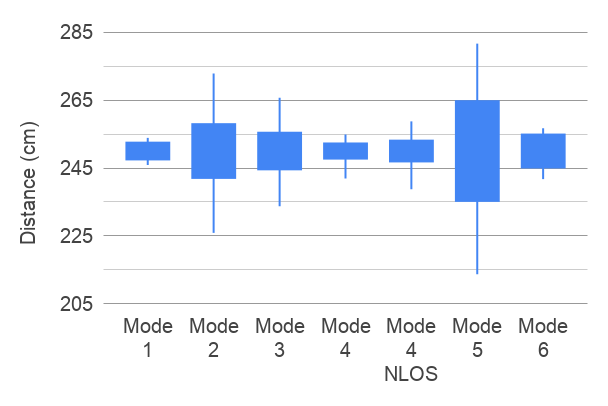} 
\end{figure}

Nevertheless, this step can be outsourced to the BlueRange gateway to save processor resources on the asset tag. 
Illustrated in Diagram \ref{fig:evabias}, the correction value for compensation of the range bias error can be determined using the equation of Friis. 
\\As a function of the average frequency and the \ac{PRF}, it approaches zero at a certain distance \cite[208]{dw1000userman}.

The graph in figure \ref{fig:evacandle} illustrates the standard (Wide bar) and maximum deviation (Narrow bar) of the experimental measurement results.
Beginning with mode 2, the use of the higher frequency \ac{PRF} improves the distance estimation around 50\%. By transmitting on channel 4 with a higher center frequency, the standard deviation is further reduced. Even in a \ac{NLOS} scenario, this mode remains advantageous compared to the other configurations. Neither the two test modes can compete with that. While the long range configuration is not reliable in the results, increasing the bandwidth allows centimeter accurate estimates as well. 
\\Nevertheless, the diagram identifies mode 4 as the most accurate one, which provides even the outliers within a close range.
The described standard deviation is important to determine the precision of the measurement. Besides the accuracy of the distance estimation, the precision is decisive for the usability of the configuration. Instead of averaging thousands of measurements, only one \ac{SS-TWR} execution is required to be sufficient in a productive system. Therefore, a low dispersion is important for the efficiency of the Asset Tracking.
\begin{figure}[ht!] 
\centering
\caption{SS-TWR - Distribution at 10m - Mode 3 \& 4}
\label{fig:evadistri3}
\includegraphics[width=3.45in]{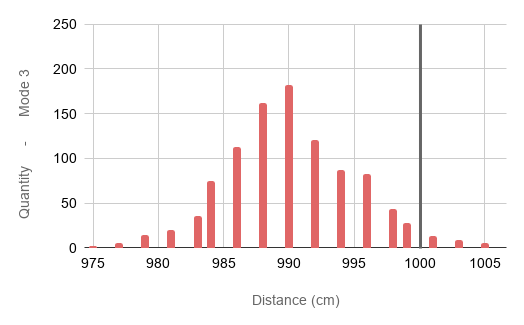}
\includegraphics[width=3.45in]{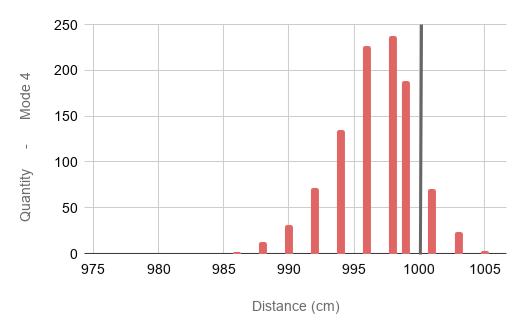}
\end{figure}
As shown in the Diagram \ref{fig:evadistri3}, the two distributions of the measurement results illustrate this advantage of mode 4.
While the estimates of the more power-saving settings are dispersed over 20 bars, the configuration 4 can restrict the remote anchors position by almost 50\%.
\begin{table}[ht!]
\centering
\caption{SS-TWR - Precision 0-30m}
\label{tab:evaprecision}
\begin{tabular}{|l|l|l|l|} 
\hline
\rowcolor[rgb]{0.855,0.91,0.988} Mode & \begin{tabular}[c]{@{}>{\cellcolor[rgb]{0.855,0.91,0.988}}l@{}}Std. Deviation\\0 - 30m \end{tabular} & \begin{tabular}[c]{@{}>{\cellcolor[rgb]{0.855,0.91,0.988}}l@{}}Std. Deviation\\0 - 10m \end{tabular} & Max. Difference \\ 
\hline
\rowcolor[rgb]{0.937,0.937,0.937} 2 & 5.94cm & 6.50cm & 30.00cm \\ 
\hline
3 & 6.48cm & 5.42cm & 71.02cm \\ 
\hline
\rowcolor[rgb]{0.937,0.937,0.937} 4 & 4.04cm & 2.44cm & 32.95cm \\ 
\hline
5 & 19.58cm & 13.55cm & 170.12cm \\ 
\hline
\rowcolor[rgb]{0.937,0.937,0.937} 6 & 4.04cm & 5.23cm & 68.11cm \\
\hline
\end{tabular}
\end{table}

According to the values in table \ref{tab:evaprecision}, the accurate measurements of configuration 4 also apply to all other measured distances.
By considering a 60\% lower total standard deviation, the settings of this mode offer clearly the highest precision for the prototype. 
In this mode, maximum and minimum deviations of less than 10cm are provided up to a distance of 15m. Especially in the range up to 7.5m, the measurement results of the configuration only deviate from the average by 2.44cm. 

Since configuration 1 is excluded due to the poor reception rate and signal quality, all remaining relevant modes have the same transmission length. 
While the settings of configuration 2 and 3 consume almost 14\% less power despite the same \ac{TWR} duration, they offer at least 100\% less precise measurements at 0-10m compared to mode 4.
Related to these results, further testing can be limited to configuration mode 4.

\subsection{Comparison of UWB \& BLE}

The selected configuration 4 was tested against an implemented replication of the existing \ac{BLE} approach in several \ac{LOS} and \ac{NLOS} setups. 
For this purpose, the range test was repeated for the determination of the \ac{BLE} distance estimation quality. For the \ac{NLOS} evaluation, the asset tag was shielded with a wooden panel (2m x 80cm x 2cm) at a distance of 20cm.

While the \ac{UWB} signals were only marginally degraded by this obstacle, the generally weaker \ac{BLE} signal experienced distraction issues, especially in the long range measurements.
Shown in table \ref{tab:evauwbrssi}, the reception rate of the \ac{BLE} distance estimates drops by another 10\% in the \ac{NLOS} scenario. From distances of 15m, almost one third of all \ac{BLE} packets are lost on the way to the anchor node.
Although, this tracking approach has the advantage that the signal only has to pass one way. 
Considering the standard deviation reveals the immense differences in the distance measurements from the \ac{UWB} prototype to the existing approach. 
Even in the test environment up to 10m, where sufficient \ac{BLE} signals are received, the measurements fluctuate widely. 
Under these optimal conditions, the average measurement can only be limited to 1m, which is 40 times less accurate than the \ac{SS-TWR}. 
In the \ac{NLOS} setup, an additional 47\% of the precision is lost.
Similarly, the measurements of the \ac{UWB} module are negatively affected with 11\% less accuracy, but the deviation is still less than 3cm. 

\begin{table}[ht!]
\centering
\caption{Distance - Comparison \ac{UWB} \& BLE 0-30m}
\label{tab:evauwbrssi}
\begin{tabular}{|l|l|l|l|l|} 
\hline
\rowcolor[rgb]{0.855,0.91,0.988} & \begin{tabular}[c]{@{}>{\cellcolor[rgb]{0.855,0.91,0.988}}l@{}}UWB \\LOS\end{tabular} & \begin{tabular}[c]{@{}>{\cellcolor[rgb]{0.855,0.91,0.988}}l@{}}UWB\\NLOS\end{tabular} & \begin{tabular}[c]{@{}>{\cellcolor[rgb]{0.855,0.91,0.988}}l@{}}BLE RSSI\\LOS\end{tabular} & \begin{tabular}[c]{@{}>{\cellcolor[rgb]{0.855,0.91,0.988}}l@{}}BLE RSSI\\NLOS\end{tabular} \\ 
\hline
\begin{tabular}[c]{@{}l@{}}Reception Rate\end{tabular} & 99.99\% & 99.98\% & 89.83\% & 79.74\% \\ 
\hline
\rowcolor[rgb]{0.937,0.937,0.937} \begin{tabular}[c]{@{}>{\cellcolor[rgb]{0.937,0.937,0.937}}l@{}}Std. Deviation\\0-30m\end{tabular} & 4.04cm & 3.27cm & \begin{tabular}[c]{@{}>{\cellcolor[rgb]{0.937,0.937,0.937}}l@{}}299.78cm \\2.6dBm\end{tabular} & \begin{tabular}[c]{@{}>{\cellcolor[rgb]{0.937,0.937,0.937}}l@{}}717.66cm \\4.54dBm\end{tabular} \\ 
\hline
\begin{tabular}[c]{@{}l@{}}Std. Deviation\\0-10m\end{tabular} & 2.44cm & 2.73cm & \begin{tabular}[c]{@{}l@{}}101.16cm \\2.45dBm\end{tabular} & \begin{tabular}[c]{@{}l@{}}148.75cm \\2.6dBm\end{tabular} \\ 
\hline
\rowcolor[rgb]{0.937,0.937,0.937} \begin{tabular}[c]{@{}>{\cellcolor[rgb]{0.937,0.937,0.937}}l@{}}Maximum \\Deviation\end{tabular} & 32.95cm & 75.46cm & \begin{tabular}[c]{@{}>{\cellcolor[rgb]{0.937,0.937,0.937}}l@{}}5657.17cm \\11.23dBm\end{tabular} & \begin{tabular}[c]{@{}>{\cellcolor[rgb]{0.937,0.937,0.937}}l@{}}4016.33cm \\19.06dBm\end{tabular} \\ 
\hline
\begin{tabular}[c]{@{}l@{}}Avg. Deviation\\250cm\end{tabular} & 11.99cm & 0.76cm & \begin{tabular}[c]{@{}l@{}}75.82cm \\1.73dBm\end{tabular} & \begin{tabular}[c]{@{}l@{}}185.86cm \\4.31dBm\end{tabular} \\ 
\hline
\rowcolor[rgb]{0.937,0.937,0.937} \begin{tabular}[c]{@{}>{\cellcolor[rgb]{0.937,0.937,0.937}}l@{}}Avg. Deviation\\750cm\end{tabular} & 5.24cm & 6.14cm & \begin{tabular}[c]{@{}>{\cellcolor[rgb]{0.937,0.937,0.937}}l@{}}241.83cm \\3.15dBm\end{tabular} & \begin{tabular}[c]{@{}>{\cellcolor[rgb]{0.937,0.937,0.937}}l@{}}699.12cm \\0.68dBm\end{tabular} \\
\hline
\end{tabular}
\end{table}

In general, the \ac{UWB} distance estimation provides excellent values up to 30m. 
In case of direct path, the measurement error does not exceed $\pm$15cm for the range of 2.5-30m (See figure \ref{fig:evabias}). Even without post-processing like bias correction, the implemented \ac{SS-TWR} provides a maximum deviation of less than 80cm for \ac{NLOS}. In contrast, the \ac{BLE} \ac{RSSI} measurements are not usable from 10m despite averaging. 
This approach, which calculates the distance based on the signal strength, is directly affected by signal attenuation.
Related to fluctuations in the meter range and outliers up to 56m, data correction is required. 
Therefore, the current BlueRange Gateway collects numerous signal strengths of the \ac{BLE} messages for Asset Tracking purpose. As a result, dynamic scenarios are less tangible.
Since the negative influences for the \ac{UWB} measurement, determining the distance over \ac{ToF}, remain within limits. Almost every measurement, which requires only 15ms, is reliable and consequently enables the detection of motion as a further advantage.

\subsection{NLOS Identification}

Since the \ac{BLE} and \ac{UWB} signals are affected by obstacles in different ways, their correlation can be used to index the absence of a direct path. While the \ac{UWB} value for 2.5m changes only by 10cm with shielding, the \ac{BLE} signal strength is more similar to the one of the 7.5m distance. With this higher spread between the \ac{UWB} measurement and \ac{RSSI} of \ac{BLE}, a \ac{NLOS} situation can be inferred. 
While the \ac{BLE} signal between two nodes cannot be used for reliable determination of \ac{NLOS} presence, the combination of both technologies is capable of this decision.

However, the direct path prediction can also be performed with the analysis of the \ac{UWB} signal strength on its own. By comparing the \ac{RSSI} of the \ac{SS-TWR} messages with the power level of the first path, \ac{LOS} can be confirmed. The implemented diagnostic interface provides the parameters for the calculation of $P_{\text{RXL}}$ and the $ P_{\text{First Path}}$ \cite[48]{dw1000userman}. 
To receive significant results, the test environment was changed to include more obstructions. Therefore, a test setup with a distance over ten meters across three rooms was chosen. While testing, the diagnostic data for three \ac{NLOS} levels was collected. In addition to the direct path, two doors with a depth of 65mm were closed and a wooden barrier with a thickness of 40mm was installed for the \ac{SS-TWR}. 
The measurements and the calculated diagnostic parameters are displayed in table \ref{tab:evanlos}.

Despite the barriers, the individual results for each parameter are quite consistent. While the \ac{BLE} values are irregularly attenuated, the difference between the $P_{\text{RXL}}$ and $ P_{\text{First Path}}$ points to a strong difference between \ac{NLOS} and \ac{LOS}. With a threshold of 7dBm, determined in an approach with 1206 measurements and an accuracy of 90\% \cite[30]{nlos}, a correct determination of the four test scenarios can be performed.

Due to the closed doors in scenario NLOS 2 and NLOS 3, the multipath signals are attenuated as well. In this case, the detection of \ac{NLOS} is not as clear as with only a blocked direct path.
\begin{table}[ht!]
\centering
\caption{\ac{UWB} NLOS Identification}
\label{tab:evanlos}
\begin{tabular}{|l|l|l|l|l|} 
\hline
\rowcolor[rgb]{0.855,0.91,0.988} & LOS & NLOS 1 & NLOS 2 & NLOS 3 \\ 
\hline
Barrier & 0cm & 6.5cm & 13cm & 17cm \\ 
\hline
\rowcolor[rgb]{0.937,0.937,0.937} Distance Est. & 981cm & 1005cm & 996cm & 1005cm \\ 
\hline
BLE RSSI & -60dBm & -64dBm & -72dBm & -76dBm \\ 
\hline
\rowcolor[rgb]{0.937,0.937,0.937} $ P_{FP}$ & 38.73dBm & 29.07dBm & 27.15dBm & 27.57dBm \\ 
\hline
$ P_{RXL}$ & 40.16dBm & 36.93dBm & 34.84dBm & 34dBm \\ 
\hline
\rowcolor[rgb]{0.937,0.937,0.937} $ P_{RXL} - P_{FP}$ & 1.43dBm & 7.86dBm & 7.69dBm & 7.23dBm \\
\hline
\end{tabular}
\end{table}

Based on this knowledge, the distance estimate can then be evaluated accordingly by the BlueRange gateway. If there are sufficient results from different anchor nodes, those with \ac{NLOS} can be prioritized lower. The detailed study of the existing positioning system shows the limitation of using only \ac{BLE} nodes to identify the direct path. In general, the system requires three nodes to locate an asset tag by calculating possible positions using the free space attenuation model and the trilateration. In this process, there is no consideration whether it is a \ac{NLOS} or \ac{LOS} situation for the individual connection. With a complex and time-consuming calibration process, the absence of a direct path can be made visible. Therefore, the signal strength of the individual node points must be measured for many reference positions and a spatial model with expected \ac{RSSI} values must be created. As a result, walls and obstacles can be visualized. By featuring high quality distance estimation as well as the detection of \ac{NLOS} occurrence, the time-consuming calibration process would become obsolete.

\subsection{Prototype Application Test}\label{sec:evaapplicationtest}
Finally, to evaluate the applicability of the prototype, the localization of an asset tag was tested in a typical office environment. For this purpose, the distance between an asset tag and two anchor nodes was estimated using the \ac{BLE} \ac{RSSI} method and the \ac{UWB} \ac{SS-TWR}. The location of the asset tag was determined by calculating the intersection of the distances. Therefore, the asset tag and the anchor nodes were kept at the same height to simplify the measurement in a two-dimensional space. 
To compare the two tracking technologies, neither for the distance bias error nor for the absence of the direct path, an error correction was made.

\begin{figure}[ht!]
\centering
\caption{UWB Indoor Localization}
\label{fig:floormap}
\includegraphics[width=3.45in]{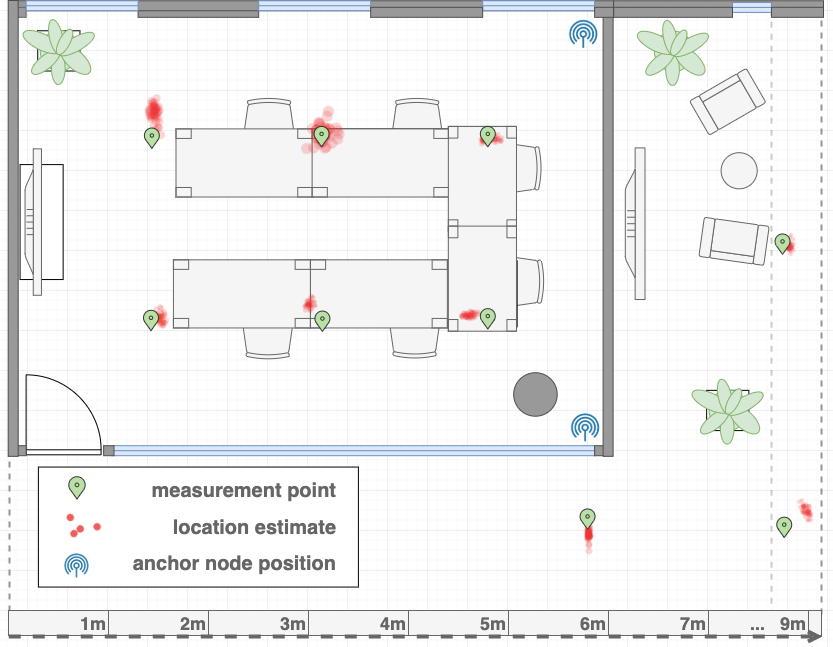}
\end{figure}

The graphic \ref{fig:floormap} is furthermore limited to the \ac{UWB} results, since the measurements based on the \ac{BLE} messages would cover the entire map with their inaccuracy. 
Overall, the accuracy and precision determined in the experiment can be confirmed in the application test. Due to the anchor being placed behind a pillar in the corner of the conference room, different typical \ac{NLOS} situations were tested. While the measurements in the conference room suffered in precision behind the column, the glass pane had only a slight negative influence on the measurements in the corridor.

A total of 11 measurements were performed, with a standard deviation of 10cm for the \ac{UWB} position determination. The locations calculated with the \ac{BLE} \ac{RSSI} could only be limited to a range of 220cm. The concrete column at point 5.50m in the graphic had the largest deflection on the \ac{UWB} signals with a standard deviation of 17.3cm. The \ac{BLE} messages lost comparatively little signal strength here.
In general, the \ac{BLE} signals vary so much that an intersection point could only be determined for 44\% of the measurements.
Whereas the positions determined with the help of \ac{UWB} are superior to those of BLE, for only one measurement at a distance of 12.5m, which was shielded by solid walls, \ac{BLE} had better reception. While 98\% of the messages were still received here, the DW1000 received only 8\% of the signals. However, even at the low reception rate, the \ac{UWB} result deviates only 76cm from the actual position, while the \ac{BLE} results are unusable.

\section{Classification of Results}\label{sec:classofresults}
The measurement results of the experimental ranging as well as the application tests reveal great accuracy and precision for the tested configuration mode 4. Nevertheless, the power consumption is 17\% higher than the most energy-efficient mode 1 and almost 80 times higher than in the basic consumption of the BlueRange node. In continuous \ac{UWB} operation, a coin cell with a capacity of 200mAh would be empty in about 3h. 
According to the \ac{RX} times in table \ref{tab:ppktime}, the asset tag's \ac{RX} operation can be started delayed. With this delay of at least 10ms the battery lifetime can be increased to almost 6.5h. The values $N_{\text{Range}}$ , $T_{\text{Offset}}$ or $F_{\text{Update}}$ of the developed concept must be adjusted to reduce the operating time of the DW1000 module to a minimum. While the simple \ac{SS-TWR} duration of just 15ms allows a detailed creation of motion profiles, the update rate $F_{\text{Update}}$ must be chosen quite low. For example, performing three rangings every five seconds increases the normal consumption by only 0.44mA. With this base consumption of 1.25mA the asset tag would only need a new battery after 160h of movement. 

This solution requires that a single measurement is sufficient to determine the distance adequately. Looking at the number of outliers and their deviation in table \ref{tab:evaprecision}, the accuracy of individual results can be restricted to a sub-meter range. As shown in the application test, with an average deviation of a few centimeters, areas of up to 200$m^2$ can be covered, depending on the installed and existing structures. 
In case that an accuracy lower than 33cm is required, measurements with a high \ac{NLOS} indication can be repeated to reduce the risk of outliers.
While the result for maximum range and the reception rate provides the reduction of the transmission power, this option for a higher energy efficiency is not further considered. Since the actual transmission represents on average only around $2\%$ of the \ac{SS-TWR} power consumption, the potential savings are marginal. In addition, the devices for production use require a housing, which increases the attenuation. Therefore, the transmission power is left at the allowed limit with maximum coverage.

Since the transmission of the \ac{UWB} message takes only about 1ms, the Single Sided \ac{TWR} can be extended for another transmission. The oscillator error correction for compensation of the clock drift can be omitted if \ac{DS-TWR} is used. Therefore, the two processing times of the asset tag and the anchor node are offset by each other. In addition, for the marginal higher amount of $0.64mA/5s$, the processing and forwarding can be moved to the anchor nodes. Subsequently, the asset tag can be put into \textit{Deep-Sleep} state earlier without the need of updating the advertising message.

\section{Conclusion}
To evaluate the benefits of the \ac{UWB} technology for BlueRange Asset Tracking as a \ac{BLE} based location estimation, the required prototype was successfully developed. 
By providing accurate distance estimation over ranges of at least 30m, the implementation matches the requirements for indoor localization.
While additional costs cannot be avoided when adding new hardware, the price of 14 euro only for the \ac{UWB} module appears, compared to the 4 euro for the FruityMesh \ac{SoC}, excessive. Furthermore, asset tags as well as anchor nodes require both a \ac{UWB} transceiver module to perform the distance estimation.
In spite of this, three times more range with precision around 4cm and measurement errors less than 15cm relativize the costs \cite{symmetryelectronics}. 

Nevertheless, the battery operation as in the BlueRange Asset Tracking would not be possible without the development of the motion-based concept to control the hardware. 
If many asset tags are in use and in motion, the update frequency of the asset tag position must be adjusted according to the desired battery life. Therefore, the most energy-efficient case here would be a single update when the asset tag is stopped.
However, accepting a high power consumption, the excellent responsiveness offers numerous other use cases. 
With a time of 15ms per measurement, accurate motion profiles can be created for fast objects. Due to various methods for minimizing interference, such as time hopping or the selection of four different preamble codes, the number of nodes stays scalable despite high message volumes \cite[269]{bensky}. 
All in all, the indoor localization based on the \ac{UWB} technology offers a promising applicability and improvement for the existing BlueRange Asset Tracking.

After the positive conclusion, the integration of the \ac{UWB} technology into the BlueRange network can be put into practice. Generally, the prototype provides a solid basis for a productive system. 
Nevertheless, the \ac{UWB} integration should be included into existing long-term tests despite the positive measurement results from this work. Besides a longer monitoring period, using an increased number of \ac{UWB} nodes would provide further insights.

Despite this comparatively low price and good technical support, the DW1000 chip was released in 2011 and received its last firmware update in 2019. Along with the release of new \ac{UWB} hardware, such as the NXP chip, new opportunities arise to reduce power consumption for high performance measurements. While the DW1000 chip is still based on the outdated IEEE 802.15.4-2011, the 2015 extension of the standard and the new version 802.15.4z offer many improvements for ranging. In particular, the introduction of security features such as encryption and improved energy efficiency through the use of techniques such as \ac{LRP} will make UWB technology even more attractive in the future \cite{ieeeuwbstandard}\cite[44]{dw1000datasheet}\cite{nxp_2020}.
\bibliographystyle{IEEEtran}
\bibliography{IEEEabrv,references}
\begin{acronym}[GFSdddddK]
\acrodef{AO}{Always On}
\acrodef{RTLS}{Real-Time Locating System}
\acrodef{BLE}{Bluetooth Low Energy}
\acrodef{GPIO}{General Purpose Input Output}
\acrodef{CW}{Continuous Wave}
\acrodef{Hz}{Hertz}
\acrodef{IoT}{Internet of Things}
\acrodef{IR-UWB}{Impulse Radio-\ac{UWB}}
\acrodef{EIRP}{Equivalent Isotropic Radiated Power}
\acrodef{BNetzA}{German Federal Network Agency}
\acrodef{LDC}{Low Duty Cycle}
\acrodef{LL}{Link Layer}
\acrodef{FCC}{Federal Communications Commission}
\acrodef{OFDM}{Orthogonal Frequency-Division Multiplexing}
\acrodef{API}{Application Programming Interface}
\acrodef{MQTT}{Message Queuing Telemetry Transport}
\acrodef{SMTP}{Simple Mail Transfer Protocol}
\acrodef{MPC}{Multipath Component}
\acrodef{IEEE}{Institute of Electrical and Electronics Engineers}
\acrodef{LAN}{Local Area Network}
\acrodef{WLAN}{Wireless Local Area Network}
\acrodef{LOS}{Line-of-Sight}
\acrodef{NLOS}{Non-Line-of-Sight}
\acrodef{TH}{Time Hopping}
\acrodef{IR}{Infrared}
\acrodef{ISM}{Industrial, Scientific and Medical}
\acrodef{SDK}{Software Development Kit}
\acrodef{SPI}{Serial Peripheral Interface}
\acrodef{SoC}{System-on-Chips}
\acrodef{GPS}{Global Positioning System}
\acrodef{GNSS}{Global Navigation Satellite Systems}
\acrodef{RSP}{Received Signal Phase}
\acrodef{BPM}{Bi-Phase Modulation}
\acrodef{RSSI}{Received Signal Strength Indication}
\acrodef{ToA}{Time of Arrival}
\acrodef{TDoA}{Time Difference of Arrival}
\acrodef{RFID}{Radio-Frequency Identification}
\acrodef{UHF}{Ultra-High-Frequency}
\acrodef{PoA}{Phase of Arrival}
\acrodef{PAC}{Preamble Acquisition Chunk}
\acrodef{PDoA}{Phase Difference of Arrival}
\acrodef{AoA}{Angle of Arrival}
\acrodef{IC}{Integrated Circuit}
\acrodef{ToF}{Time of Flight}
\acrodef{LoRaWAN}{Long Range Wide Area Network}
\acrodef{SS-TWR}{Single-Sided Two-Way Ranging}
\acrodef{PSM}{Pulse Shape Modulation}
\acrodef{OOK}{On-Off Keying}
\acrodef{PPM}{Pulse Position Modulation}
\acrodef{NFC}{Near Field Communication}
\acrodef{UART}{Universal Asynchronous Receiver Transmitter}
\acrodef{USB}{Universal Serial Bus}
\acrodef{UUID}{Universally Unique Identifier}
\acrodef{UWB}{Ultra-Wideband}
\acrodef{SMD}{Surface-Mounted Device}
\acrodef{PHY}{Physical Layer}
\acrodef{PHR}{\ac{PHY} header}
\acrodef{OTP}{One-Time Programmable}
\acrodef{LO}{Local Oscillator}
\acrodef{DS-TWR}{Double-Sided Two-Way Ranging}
\acrodef{CIR)}{Channel Impulse Response}
\acrodef{TCXO} {Temperature Compensated Crystal Oscillator}
\acrodef{PRF}{Pulse Repetition Frequency}
\acrodef{FSPL}{Free-Space Path Loss}
\acrodef{RTT}{Real-Time Transfer}
\acrodef{FM}{Frequency Modulation}
\acrodef{PDP}{Power Delay Profile}
\acrodef{SHR}{Synchronization Header}
\acrodef{SFD}{Start of Frame Delimiter}
\acrodef{BPSK}{Binary Phase Shift Keying}
\acrodef{PAM}{Pulse Amplitude Modulation}
\acrodef{FEC}{Forward Error Correction}
\acrodef{SECDED}{Single-Error-Correct Double-Error-Detect}
\acrodef{RS}{Reed Solomon}
\acrodef{PSR}{Preamble Symbol Repetition}
\acrodef{PSDU}{Physical Layer Service Data Unit}
\acrodef{BPM-BPSK}{Burst Position Modulation and Binary Phase Shift Keying}
\acrodef{TX}{Transmit}
\acrodef{RX}{Receive}
\acrodef{FP}{First Path}
\acrodef{TWR}{Two-Way Ranging}
\acrodef{FCS}{Frame Check Sequence}
\acrodef{CIR}{Channel Impulse Response}
\acrodef{PPK}{Power Profiler Kit}
\acrodef{LRP}{Low Rate Pulse Repetition}	
\acrodef{SNR}{Signal-to-Noise Ratio}
\acrodef{sonar}{sound navigation and ranging}
\end{acronym} 
\end{document}